\documentstyle[epsf,times]{jaa}

\begin{document}
\title[LOFAR Results and Prospects]{LOFAR: Recent imaging results \& future prospects} 
\author[G. Heald et al.]%
       {George Heald,$^{1}$\thanks{e-mail: heald@astron.nl} Michael R. Bell,$^2$ Andreas Horneffer,$^3$
       \newauthor
       Andr\'e R. Offringa,$^4$ Roberto Pizzo,$^1$ 
       \newauthor
       Sebastiaan van der Tol,$^5$ Reinout J. van Weeren,$^5$ 
       \newauthor
       Joris E. van Zwieten,$^1$ James M. Anderson,$^3$ Rainer Beck,$^3$
       \newauthor
       Ilse van Bemmel,$^1$ Laura B\^irzan,$^5$ Annalisa Bonafede,$^6$
       \newauthor
       John Conway,$^7$ Chiara Ferrari,$^8$ Francesco De Gasperin,$^2$
       \newauthor
       Marijke Haverkorn,$^{1,5}$ Neal Jackson,$^9$ Giulia Macario,$^{10}$ 
       \newauthor
       John McKean,$^1$ Halime Miraghaei,$^{11,3,12}$ Emanuela Orr\`u,$^{13}$
       \newauthor
       David Rafferty,$^5$ Huub R\"ottgering,$^5$ Anna Scaife,$^{14}$
       \newauthor
       Aleksandar Shulevski,$^4$ Carlos Sotomayor,$^{15}$ Cyril Tasse,$^{16}$
       \newauthor
       Monica Trasatti,$^{17}$ Olaf Wucknitz,$^{17}$
       \newauthor
       on behalf of the LOFAR collaboration \\ 
       $^1$ASTRON, Postbus 2, 7990 AA Dwingeloo, the Netherlands\\
       $^2$Max-Planck-Institut f\"ur Astrophysik, Karl-Schwarzschildstra{\ss}e 1,\\
       85741 Garching, Germany\\
       $^3$Max-Planck-Institut f\"ur Radioastronomie, Auf dem H\"ugel 69, 53121 Bonn, Germany\\
       $^4$Kapteyn Astronomical Institute, University of Groningen, PO Box 800,\\
       9700 AV Groningen, the Netherlands\\
       $^5$Leiden Observatory, Leiden University, PO Box 9513, 2300 RA Leiden,\\
       the Netherlands\\
       $^6$Jacobs University Bremen, Campus Ring 1, 28759 Bremen, Germany\\
       $^7$Onsala Space Observatory, Dept. of Earth and Space Sciences,\\
       Chalmers University of Technology, SE-43992 Onsala, Sweden\\
       $^8$UNS, CNRS UMR 6202 Cassiop\'ee, Observatoire de la C\^ote d'Azur, Nice, France\\
       $^9$Jodrell Bank Centre for Astrophysics, School of Physics and Astronomy,\\
       The University of Manchester, Oxford Road, Manchester M13 9PL, United Kingdom\\
       $^{10}$INAF - Istituto di Radioastronomia, via Gobetti 101, 40129 Bologna, Italy\\
       $^{11}$Institute for Studies in Theoretical Physics and Mathematics, Tehran, Iran\\
       $^{12}$Sharif University of Technology, Tehran, Iran\\
       $^{13}$Radboud University Nijmegen, Heijendaalseweg 135, 6525 AJ Nijmegen,\\
       the Netherlands\\
       $^{14}$Dublin Institute for Advanced Studies, 31 Fitzwilliam Place, Dublin 2, Ireland\\
       $^{15}$Astronomisches Institut der Ruhr-Universit\"at Bochum,\\
       Universit\"atsstr. 150, 44780, Bochum, Germany\\
       $^{16}$GEPI, Observatoire de Paris-Meudon, 5 place Jules Janssen, 92190 Meudon, France\\
       $^{17}$Argelander-Institut f\"ur Astronomie, Auf dem H\"ugel 71, 53121 Bonn, Germany}

\maketitle
\label{firstpage}
\begin{abstract}
The Low Frequency Array (LOFAR) is under construction in the Netherlands and in several surrounding European countries. In this contribution, we describe the layout and design of the telescope, with a particular emphasis on the imaging characteristics of the array when used in its ``standard imaging'' mode. After briefly reviewing the calibration and imaging software used for LOFAR image processing, we show some recent results from the ongoing imaging commissioning efforts. We conclude by summarizing future prospects for the use of LOFAR in observing the little-explored low frequency Universe.
\end{abstract}


\epsfclipon

\section{Introduction and LOFAR Status}\label{section:intro}

The Low Frequency Array (LOFAR) is still being constructed, but is already producing high-quality interferometric data on baselines ranging from about 100~m, up to more than 1000~km \nocite{wucknitz_2010} (Wucknitz 2010), between frequencies as low as 10~MHz up to 240~MHz. The telescope is highly flexible, and offers many opportunities for investigations of (among many other things) diffuse emission from relativistic plasmas. LOFAR is unique in enabling arcsecond imaging at such low frequencies.

The primary scientific drivers for LOFAR are encapsulated within the so-called Key Science Projects (KSPs), which are Surveys (see R\"ottgering et al., this volume), Cosmic Magnetism, Epoch of Reionization, Transients, Cosmic Rays, and Solar Science and Space Weather. The first two are the projects which are most relevant to these proceedings. They include observations of the Milky Way, nearby galaxies, and galaxy clusters, among diverse other targets.

LOFAR has been described elsewhere in detail (see, e.g., \nocite{stappers_etal_2011} Stappers et al. 2011). Here, we briefly summarize the state of the station rollout.\footnote{For continued up-to-date information on the rollout of the array, the reader is referred to {\tt http://www.astron.nl/$\sim$heald/lofarStatusMap.html}} The array is built up out of a vast number of simple dipoles, grouped in clusters called stations. The dipoles and stations are designed differently for the Low Band Array (LBA; 10--80 MHz) and the High Band Array (HBA; 110--240 MHz). As of early May 2011, there are 24 core stations (within about 2~km of the center of the array near the small village of Exloo in the Netherlands), 7 remote stations (within about 100~km), and 8 international stations in France, Germany, Sweden, and the UK (see Figure 1). An important feature of the station design is that the HBA dipoles are split into two substations in the core; these substations can (optionally) be correlated separately for increased sensitivity to sources with large angular size.

Correlation is performed by a BlueGene supercomputer in the city of Groningen. Post-processing is handled by a number of different software pipelines which are currently under heavy development. The software aspect of the LOFAR system is of crucial importance. The pipeline which performs processing of the imaging data is the LOFAR Standard Imaging Pipeline, which has been described by \nocite{heald_etal_2010} Heald et al. (2010) and is summarized briefly in Section 3. 

\begin{figure}
\centering
\epsfxsize=0.32\textwidth \epsfbox{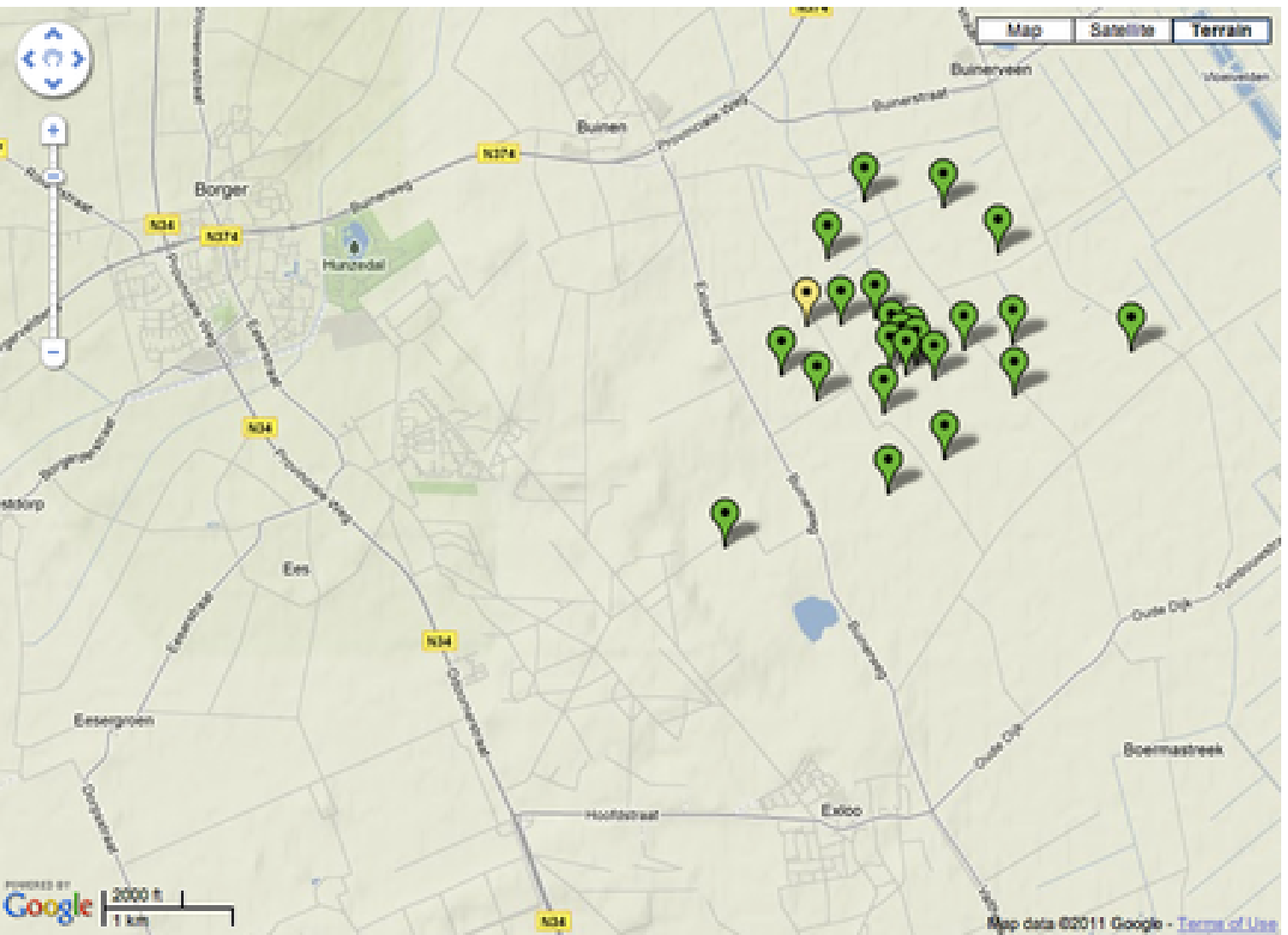}\hfill{}
\epsfxsize=0.32\textwidth \epsfbox{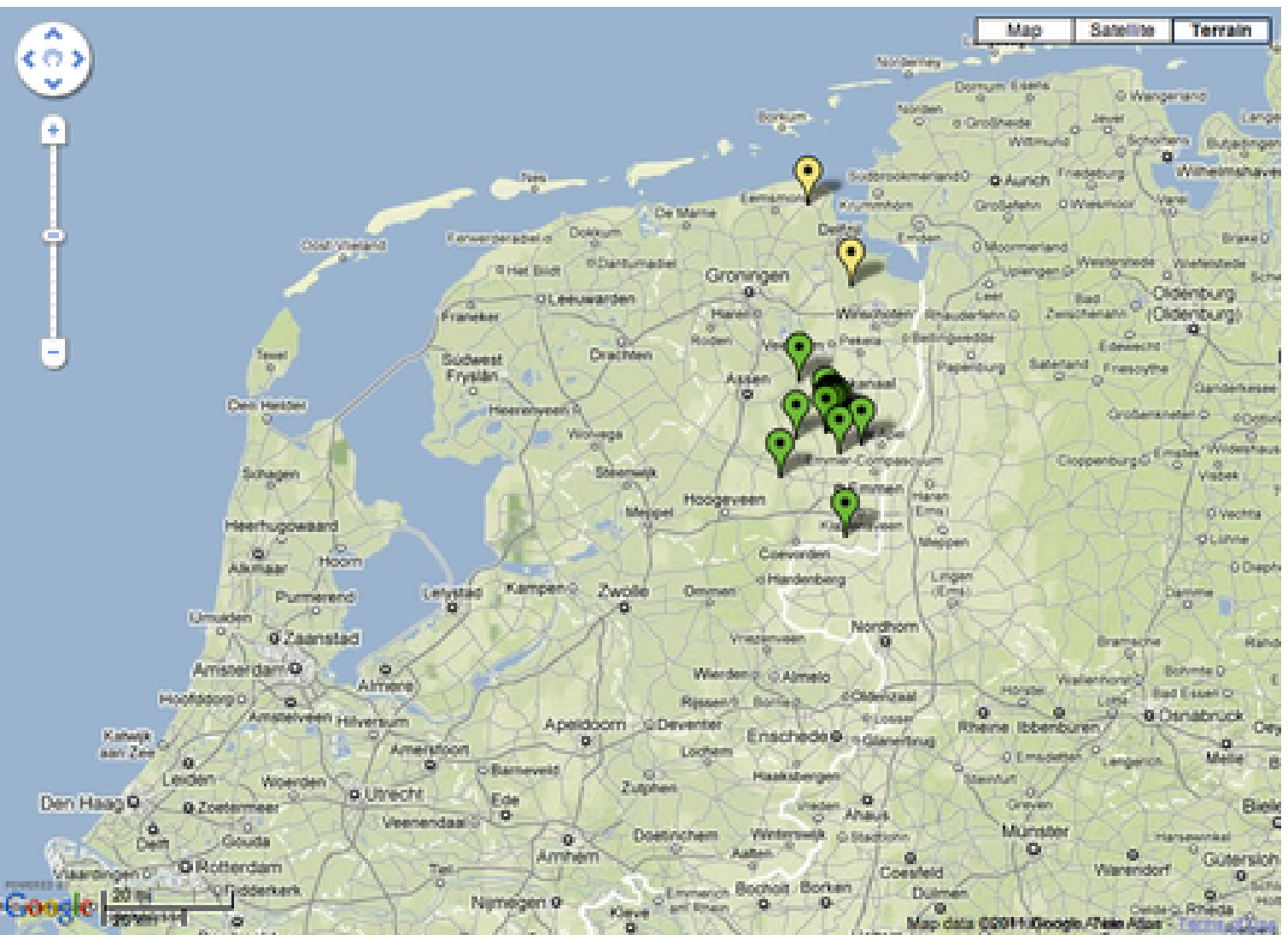}\hfill{}
\epsfxsize=0.32\textwidth \epsfbox{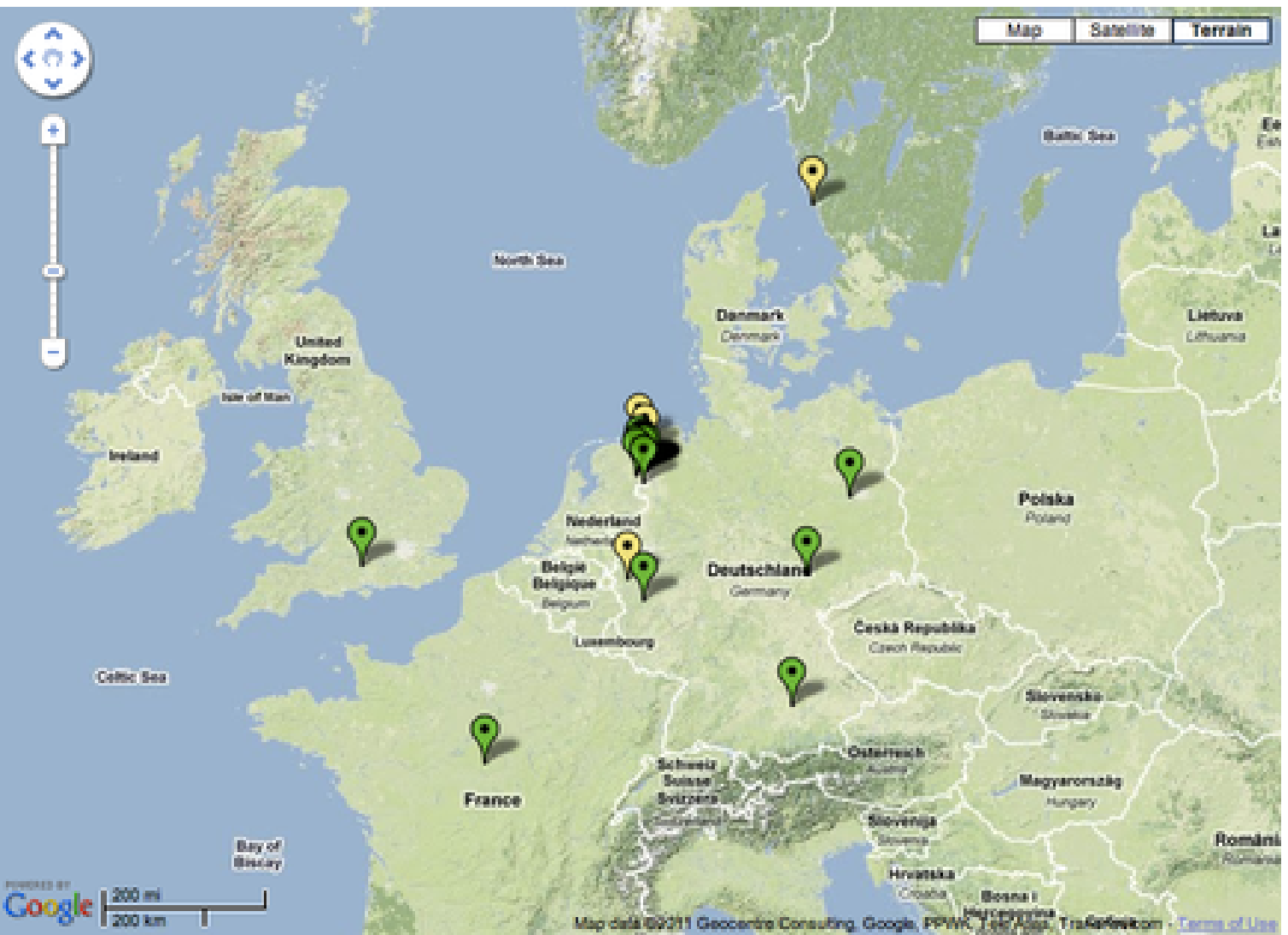}
\caption{Existing LOFAR layout on three different scales: at the LOFAR core ({\it left}; the image is $\approx$10~km wide), remote baselines ({\it middle}; the image is $\approx$320~km wide), and international scale ({\it right}; the image is $\approx$2500~km wide). Green markers indicate stations which are fully completed and validated for normal use in the array; yellow markers indicate stations which are partially completed. Seven Dutch remote stations, for which significant construction work has not yet started (as of early May 2011), are not displayed. The images are from the LOFAR Status Map, which uses Google Maps.}
\end{figure}

\section{Imaging Capabilities}\label{section:imaging}

In this section, we describe the sensitivity and resolution of the completed LOFAR array (24 core + 16 remote + 8 international = 48 stations), and address the tradeoffs between these critical parameters. For imaging observations of extended objects, the best compromise will have to be identified by selecting a particular $uv$ weighting and tapering scheme.

To illustrate the $uv$ coverage of the full array, we generated simulated data sets using the locations of all 48 current and future stations. For sources at a particular representative declination of $\delta=+48^\circ$ (corresponding to the coordinates of the bright calibrator source 3C196), we show the resulting $uv$ coverage at different linear scales in Figure 2. Note that the fractional bandwidth can be extremely high (up to 48 MHz of total bandwidth is available, and can be distributed over larger frequency spans) and this leads to significantly greater filling of the $uv$ plane (in the radial direction). This is shown for a particular case in Figure 3. Longer integration times fill the $uv$ plane azimuthally. The synthesized beams corresponding to these $uv$ coverage scenarios are illustrated by van Haarlem et al. (2011, in prep.).

\begin{figure}
\centering
\epsfxsize=0.32\textwidth \epsfbox{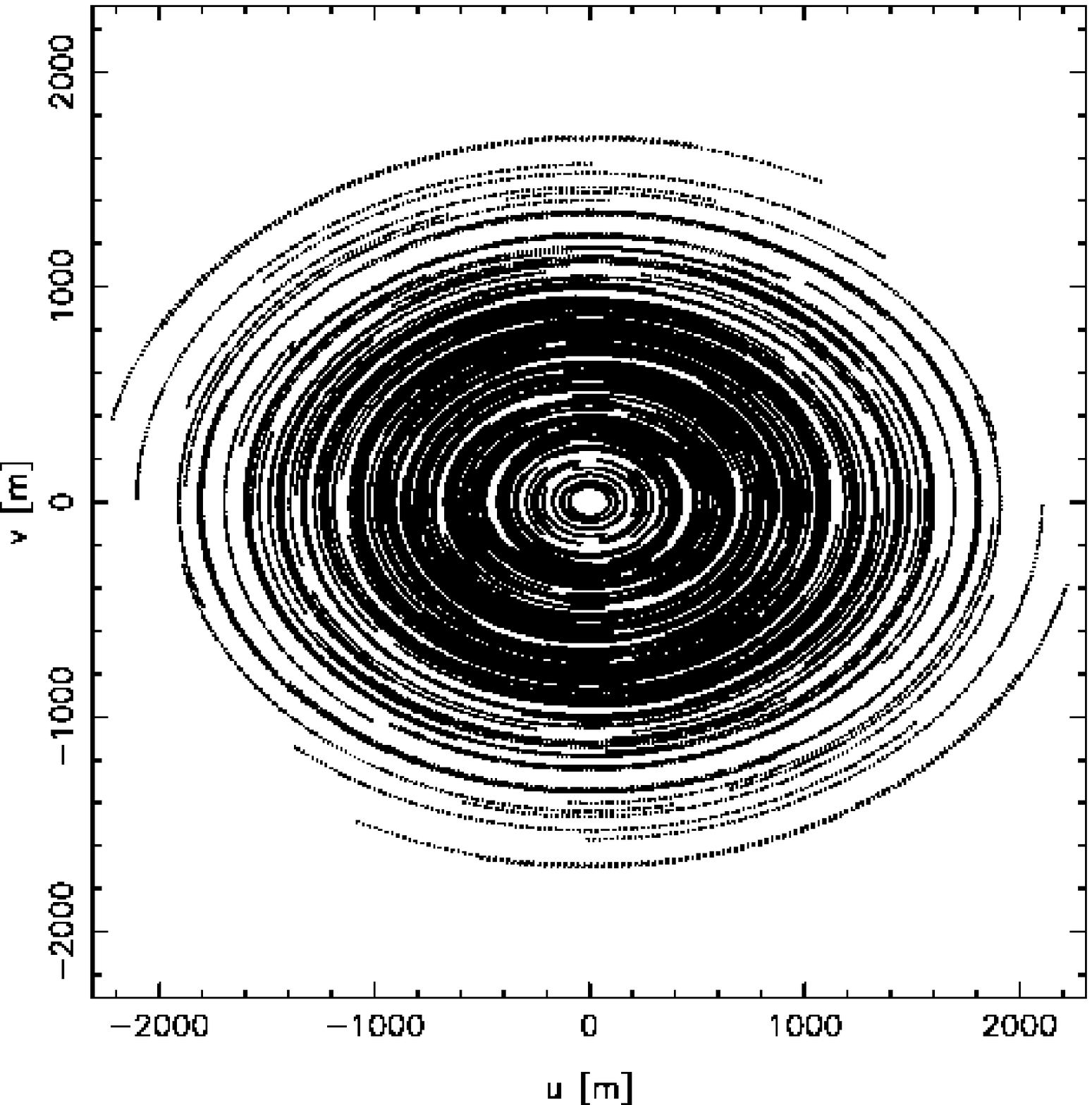}\hfill{}
\epsfxsize=0.32\textwidth \epsfbox{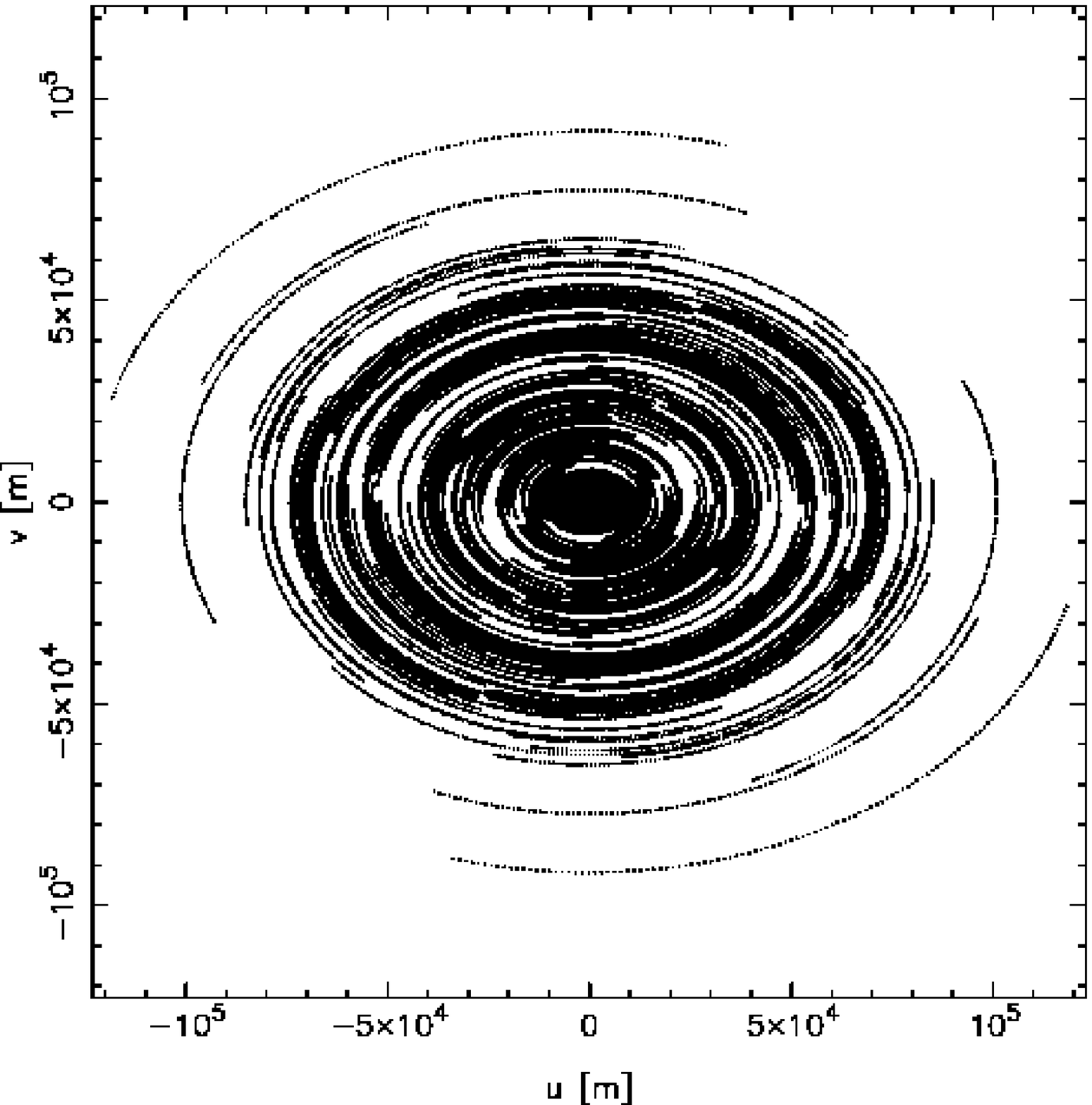}\hfill{}
\epsfxsize=0.32\textwidth \epsfbox{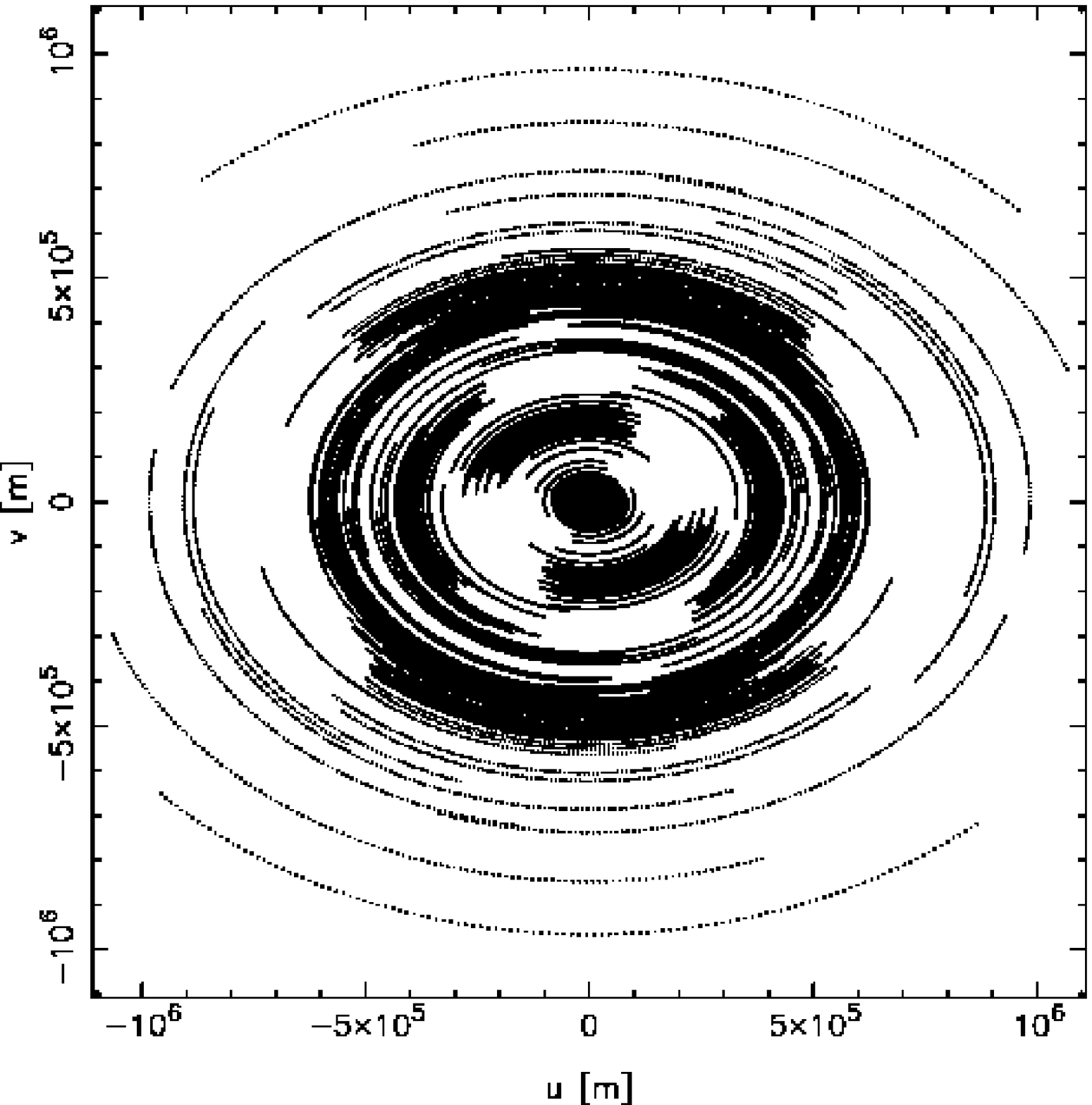}
\caption{Simulated single-frequency, long track (6 hours) $uv$ coverage for the full LOFAR array. The source crosses meridian near the middle of the observation. Left: core stations only (axis ranges $\sim\pm$2.3~km); middle: core and Dutch remote stations ($\sim\pm$120~km); right: all stations ($\sim\pm$1100~km). One point is plotted every 3 minutes. The $uv$ coverage is plotted for LBA observations; the situation for HBA observations is nearly identical if the core stations are not used in their ``split'' mode.}
\end{figure}

\begin{figure}
\centering
\epsfxsize=0.32\textwidth \epsfbox{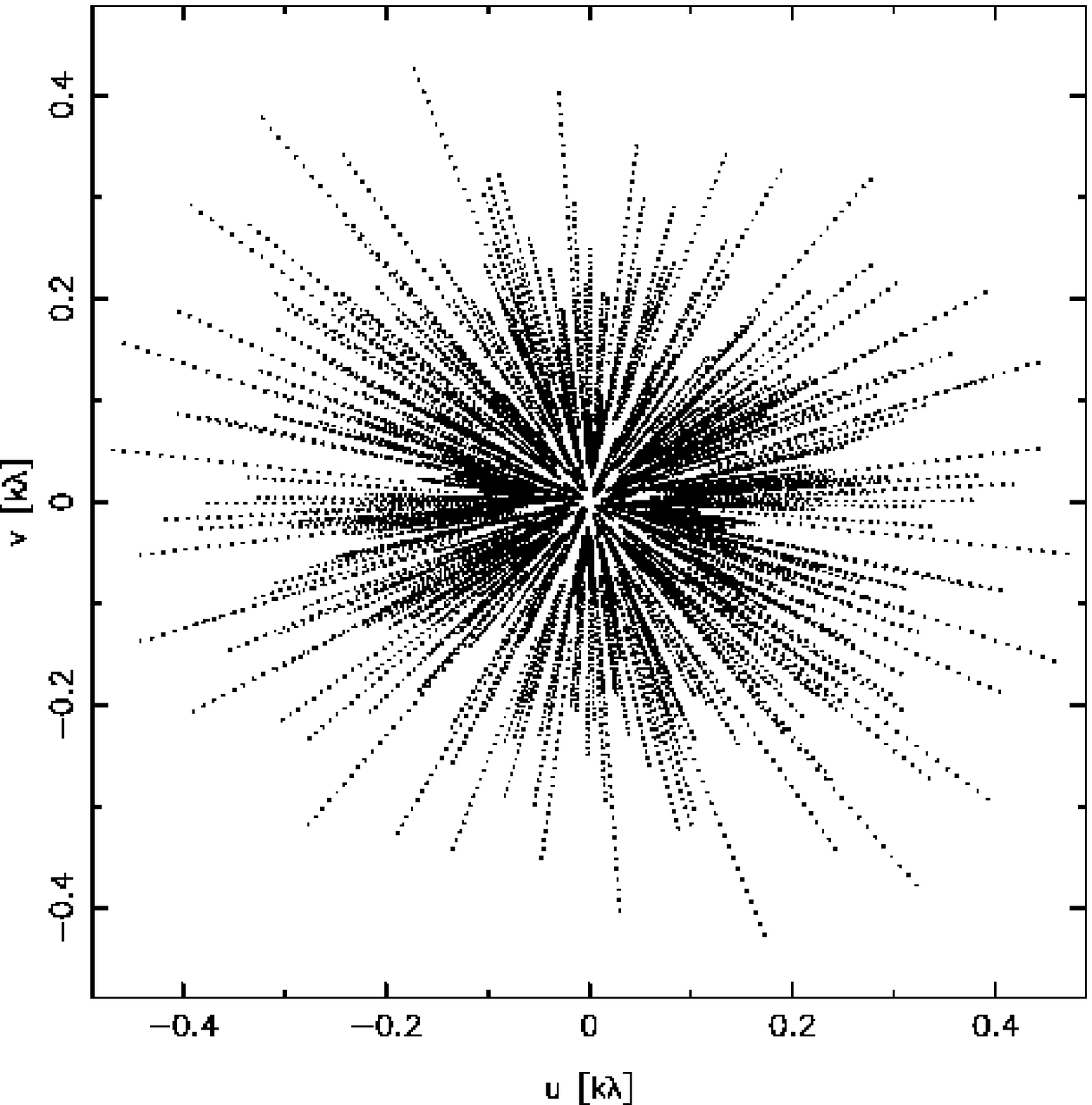}\hfill{}
\epsfxsize=0.32\textwidth \epsfbox{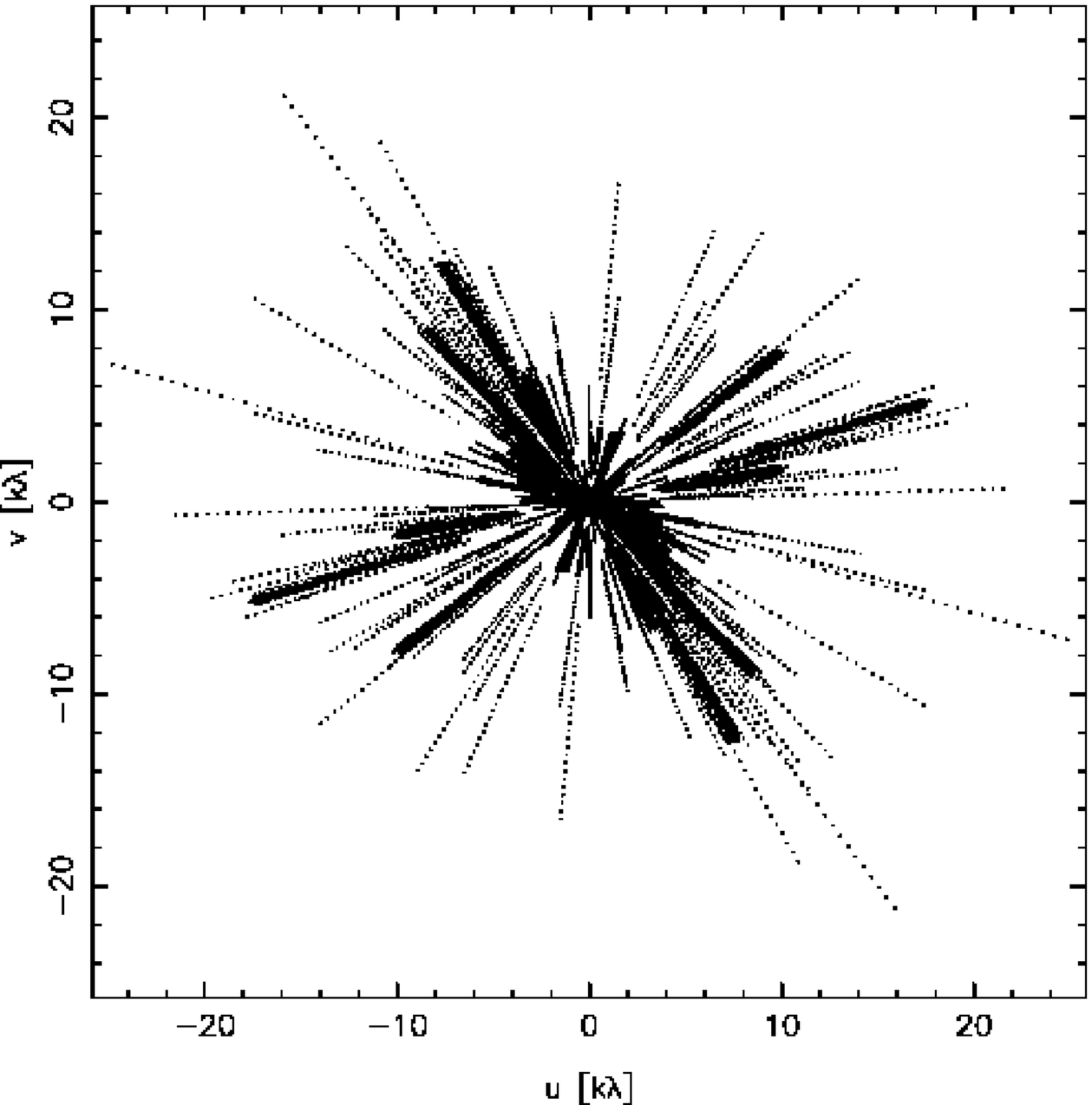}\hfill{}
\epsfxsize=0.32\textwidth \epsfbox{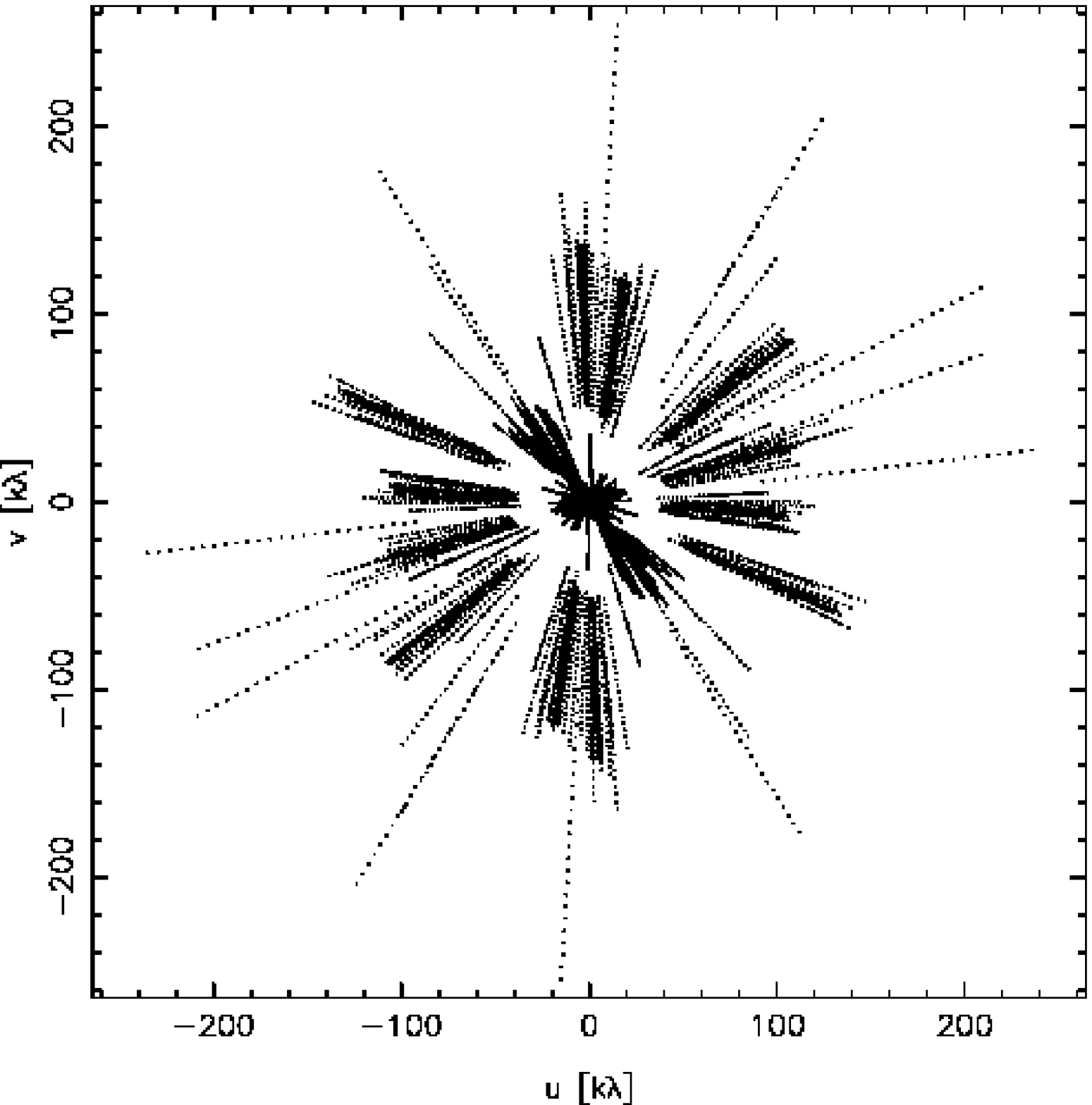}
\caption{Simulated broadband, snapshot $uv$ coverage for the full LOFAR array. Left: core only  (axis ranges $\sim\pm0.5\,\mathrm{k}\lambda$); middle: core and Dutch remote stations ($\sim\pm26\,\mathrm{k}\lambda$); right: all stations ($\sim\pm260\,\mathrm{k}\lambda$). One point is plotted every 2 MHz (corresponding to 10 subbands), in the range 30--78 MHz.}
\end{figure}

It is of interest to investigate the sensitivity of LOFAR to extended emission. We therefore next consider the sensitivity of the array at various angular scales, using frequencies of 60 MHz and 150 MHz. For this, we assume the (theoretically calculated) System Equivalent Flux Densities (SEFDs) given by van Haarlem et al. (2011, in prep.). At 60 MHz (using the {\tt LBA\_OUTER} station mode), these are 29.59 kJy for core and remote stations and 14.76 kJy for international stations; at 150 MHz, 2.82 kJy for (each half of) core stations, 1.41 kJy for remote stations, and 0.71 kJy for international stations. Simulated observations were produced using a total observing length of one hour (near transit), a bandwidth of 0.2 MHz (corresponding to a single subband, of which 244 are available in a single LOFAR observation), and the same declination as was used for the $uv$ plots in Figures 2 and 3. Gaussian noise with statistics appropriate to the frequencies used, the product of bandwidth and integration time, and the station types, was put into the simulated data sets. These artificial, noise-only, visibility data sets were imaged using the {\tt CASA} imager (task {\tt clean}), using either natural or uniform weighting, together with a wide range of (outer) $uv$ taper values. Finally, the rms of each output image was calculated. The results are shown as a function of beam size in Figure 4 (here, the effective synthesized beam radius means the radius of a circular beam with the same angular area as the actual, somewhat elliptical, synthesized beam).

Although the SEFDs used for these simulations are theoretical, currently existing data is now being used to test their validity. By observing a calibrator of known flux density, and inspecting the signal-to-noise ratio of the visibilities, a direct measure of the SEFD is obtained. Early estimates indicate that for the HBA, the SEFD estimates are, if anything, somewhat pessimistic, but further analysis is needed to confirm this.

It should also be cautioned that the actual measured noise in LOFAR images is likely to be dominated by calibration and imaging errors, certainly in the early years of LOFAR operation. The noise values given here are appropriate in case of perfect calibration and imaging.

\begin{figure}
\centering
\epsfxsize=0.49\textwidth \epsfbox{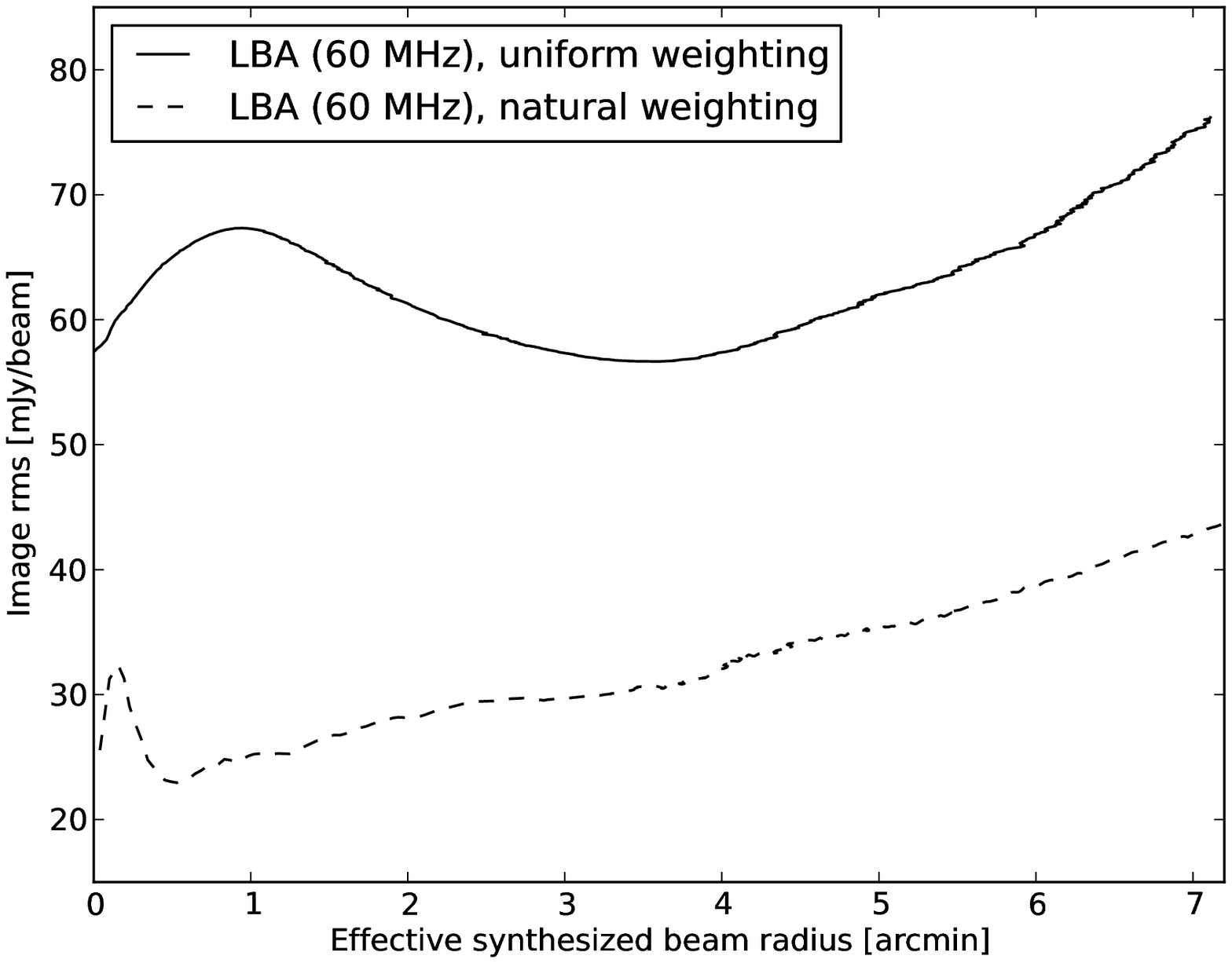}\hfill{}
\epsfxsize=0.49\textwidth \epsfbox{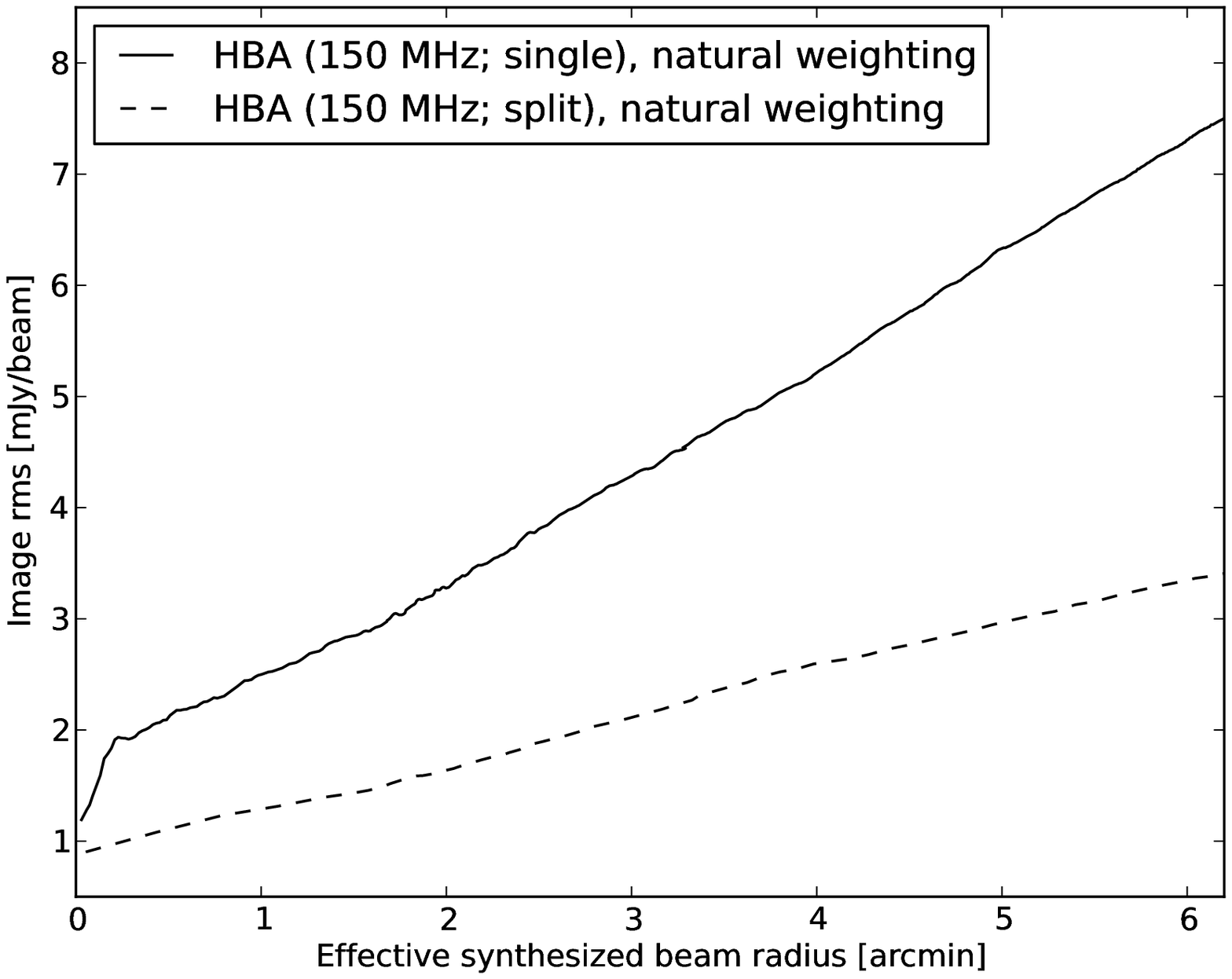}
\caption{LOFAR image noise estimates, as a function of resolution, for various combinations of visibility weighting and tapering. All current and planned stations are included (though the international stations contribute little on the large angular scales shown here). Left: LBA (60 MHz); right: HBA (150 MHz). The LBA sensitivities are shown for both uniform and natural weighting. The HBA sensitivities are shown for natural weighting only. The simulated data set included 1 hour observing time and covered one subband (200 kHz). Only theoretical sensitivities are taken into account; calibration errors will increase the noise values in actual LOFAR images.}
\end{figure}

\section{The LOFAR Standard Imaging Pipeline}

In this section we describe the LOFAR Standard Imaging Pipeline, highlight the key components, and illustrate their performance on recent imaging data. A more complete description of the pipeline is given by \nocite{heald_etal_2010} Heald et al. (2010), and will be presented in full in a forthcoming paper. 

The pipeline framework is shown by \nocite{heald_etal_2010} Heald et al. (2010; their Figure 1). The most important components are (i) the flagger and data compression utility, called the ``Default Pre-Processing Pipeline'' or DPPP (labeled DP3 in the pipeline figure); (ii) the calibration engine, called BlackBoard Selfcal (BBS); (iii) the imager; and (iv) the sky model database. Each of these are briefly discussed in turn.

Flagging of radio frequency interference (RFI) is of crucial importance. Despite the relatively high level of RFI in northern Europe, excellent rejection without significant loss of data is possible thanks to the high frequency and time resolution of LOFAR data (recent observations use 4 kHz channels and 1--3 second integrations, depending on observing frequency). Typically, $<$10\% of data are lost due to RFI flagging, and at many frequencies the statistics are even better. See Figure 5 for representative examples. In this Figure, the flagging has been done with DPPP, using the algorithm described by \nocite{offringa_etal_2010} Offringa et al. (2010).

\begin{figure}
\centering
\epsfxsize=0.49\textwidth \epsfbox{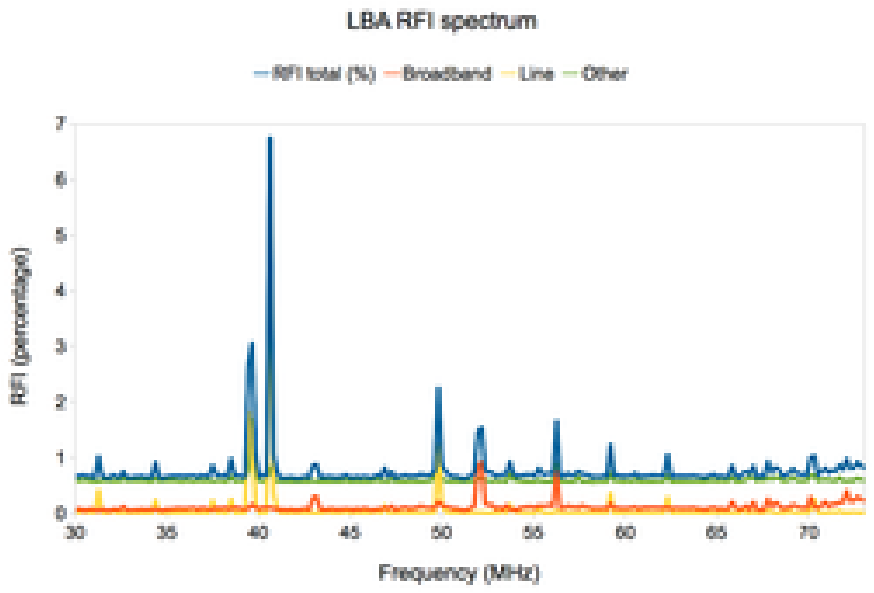}\hfill{}
\epsfxsize=0.49\textwidth \epsfbox{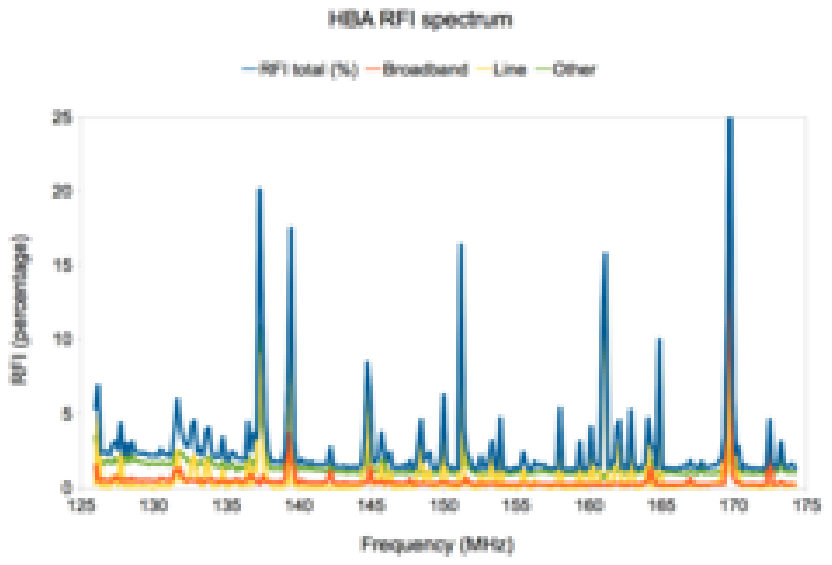}
\caption{LOFAR RFI statistics, derived using single representative observations in the indicated bands. RFI identified as broadband is show in red; narrowband RFI is shown in yellow; and RFI which is harder to classify is shown in green. The total RFI percentages are shown in blue. The LBA statistics (left) range from 30--74 MHz and 0--7\% RFI occupancy, and the HBA statistics (right) from 126--174 MHz and 0--25\% RFI occupancy.}
\end{figure}

The calibration is based on the Measurement Equation \nocite{smirnov_2011} (see, e.g., Smirnov 2011). BBS is built in such a way that the inherent direction dependence of many of the calibration problems are explicitly accounted for. This enables solving for gain solutions in multiple directions simultaneously. A particular cause for concern is the phase fluctuations induced by the Earth's ionosphere. The Sun is now becoming more active, making ionospheric disturbances stronger. Despite this, we are still able to track rapid phase fluctuations as seen in the solutions determined using BBS (see, e.g., Figure 3 of \nocite{heald_etal_2010} Heald et al. 2010). BBS will be used to apply the ionospheric modeling algorithm described by \nocite{intema_etal_2009} Intema et al. (2009).

The imaging step itself is a difficult task for LOFAR --- the nature of the dipoles, and their fixed orientation on the ground, makes the sensitivity pattern of the telescope not only a function of angular position and observing frequency, but also a (strong) function of time. One of the major consequences is that deconvolution routines such as {\tt CLEAN} do not function as expected, since the synthesized beam changes significantly as a function of position in the sky. To account for this, we are working on a LOFAR implementation of the A-projection algorithm \nocite{bhatnagar_etal_2008} (Bhatnagar et al. 2008). In the meantime, LOFAR images are limited by deconvolution errors. This is mitigated by subtracting the brightest sources in the visibility domain prior to imaging.

The LOFAR sky model is needed for calibrating the telescope in arbitrary locations on the sky. Initial models, based on extrapolations from existing radio frequency catalogs, together with previous pointed observations, are typically used as starting points in the commissioning period. An all-sky calibration survey, aiming to produce a catalog of the brightest sources in the LOFAR sky (and, importantly, their spectral behavior), will take place later in the commissioning period.

\section{Recent LOFAR imaging results}

The ongoing commissioning of LOFAR has produced some excellent imaging results. Some recent progress is highlighted here. For additional examples, see \nocite{heald_etal_2010} Heald et al. (2010) and \nocite{mckean_etal_2011} McKean et al. (2011).

LOFAR's LBA system has an extremely large field of view. The individual dipoles effectively see the entire sky, and although the station beamformer concentrates the sensitivity in a particular region of the sky, strong sources (such as Cygnus A and Cassiopeia A) still contribute significantly to the visibility function for an arbitrary observation, and in fact those two sources typically {\it dominate} the visibility function when they are above the horizon. Thus the brightest sources in the sky must always be accounted for. Brute-force strategies, solving for the gains in the directions of the brightest sources and then subtracting them from the visibilities, have been successful but are highly computationally expensive and time consuming. Recently, we have been testing and finding remarkable success with the ``demixing'' method described by \nocite{vdtol_etal_2007} van der Tol et al. (2007).

Illustrating the results from the demixing technique, we show in Figure 6 images of two 3C sources, which were imaged at a frequency of 58 MHz. To remove the effects of off-axis sources, demixing was applied using models of Cassiopeia A, Cygnus A, and Taurus A. High-resolution models based on previous observations of the 3C sources themselves were also utilized. Each demixing operation results in a number of visibility datasets, each of which nominally contain contributions from only one of the sources included in the sky model. We have found that not only the target source can be well calibrated and imaged following demixing, but also the off-axis bright sources as well --- in one recent test, Cassiopeia A was successfully imaged (not shown), despite being located some 127 degrees away from the target coordinates.

\begin{figure}
\centering
\epsfxsize=0.49\textwidth \epsfbox{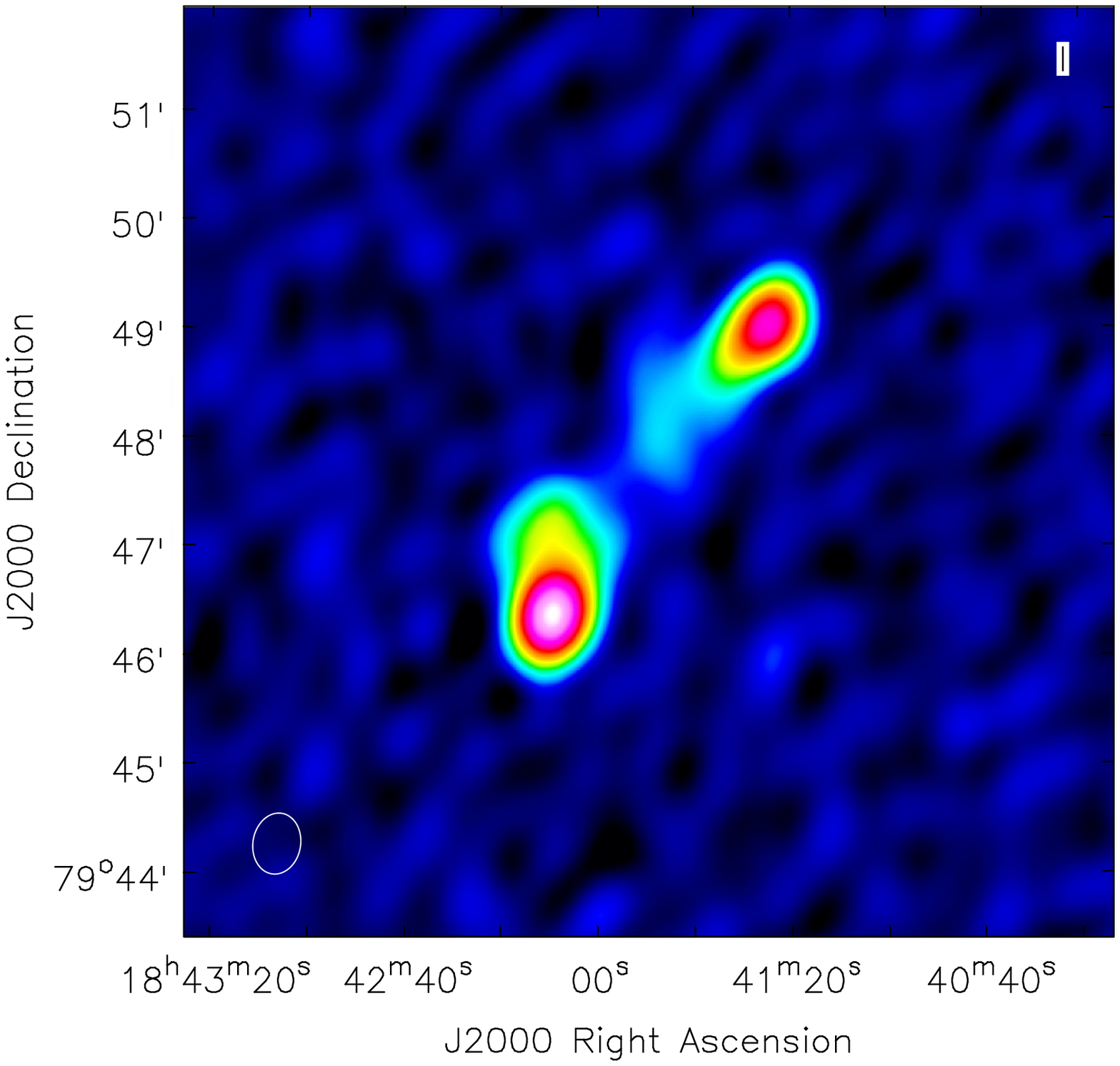}\hfill{}
\epsfxsize=0.49\textwidth \epsfbox{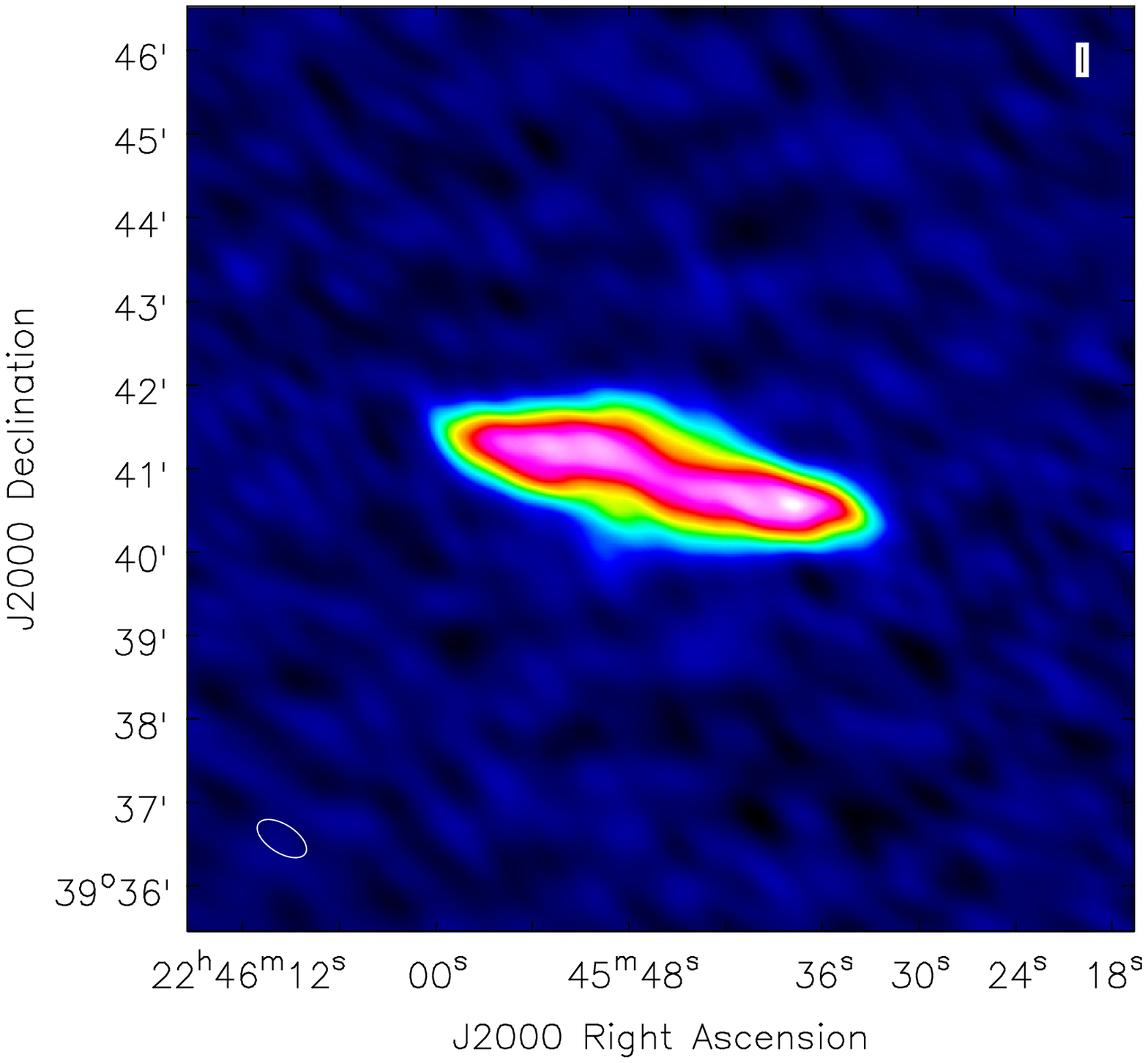}
\caption{LOFAR images of sources calibrated after ``demixing'' was applied. On the left, a 58 MHz image of 3C390.3 is shown. The noise level in the image is approximately 270~mJy~beam$^{-1}$, with a synthesized beam size of $26^{\prime\prime}\times34^{\prime\prime}$, resulting from uniform weighting. On the right, an image of 3C452 at the same frequency is displayed. The noise level is approximately 220~mJy~beam$^{-1}$, with a synthesized beam size of $40^{\prime\prime}\times20^{\prime\prime}$.}
\end{figure}

Polarized low-frequency radio emission has also been confidently detected with LOFAR. To illustrate this, we present recent results from the field of 3C66 in Figure 7. 3C66A and 3C66B themselves are shown in total intensity contours, toward the northeast corner of the frames. About a degree to the southwest is a millisecond pulsar, J0218+4232 (not shown in the total intensity contours). This pulsar is known to have a rotation measure of about -61~rad~m$^{-2}$, from previous Westerbork Synthesis Radio Telescope (WSRT) observations \nocite{navarro_etal_1995} (Navarro et al. 1995). LOFAR HBA data were imaged in Stokes Q and U in 800 channels over a frequency span of 50 MHz. Using the Rotation Measure (RM) Synthesis technique \nocite{brentjens_debruyn_2005} (Brentjens \& de Bruyn 2005), the groups of polarization channel maps were converted to an image cube, where the third axis is a coordinate called Faraday depth (which, under simple circumstances, is equivalent to rotation measure). Three frames from this cube are shown in Figure 7. Polarized emission is clearly visible from the pulsar at the correct Faraday depth (apart from its sign, which is incorrect because of a software mismatch in coordinate definitions). The bright arc-shaped features to the northwest and southeast of the pulsar itself are synthesized beam sidelobes (the polarization images were not deconvolved using e.g. {\tt CLEAN}). The empty frame (at -100~rad~m$^{-2}$) gives an indication of the noise level in the polarization images; the polarized emission from the pulsar is detected with a signal-to-noise ratio of about 25. The frame at 0~rad~m$^{-2}$ illustrates that there is a good deal of instrumental polarization which has not yet been accounted for in the calibration. The frequency dependence of the instrumental polarization means that it appears only near 0~rad~m$^{-2}$.

Since the Faraday depth is a function of the electron density and magnetic field strength in the medium through which the radio waves propagate, and the Earth's ionosphere is both magnetized and ionized, changes in the Faraday depth of polarized celestial sources track changes in the ionosphere --- particularly the electron content, which varies on short timescales. Using the observations shown here, the Faraday depth of the pulsar was tracked over the course of about 8 hours. The results showed that the Faraday depth remained stable on 1 hour timescales, indicating that there were no detectable large scale fluctuations in the ionosphere throughout that particular night.

\begin{figure}
\centering
\epsfxsize=0.49\textwidth \epsfbox{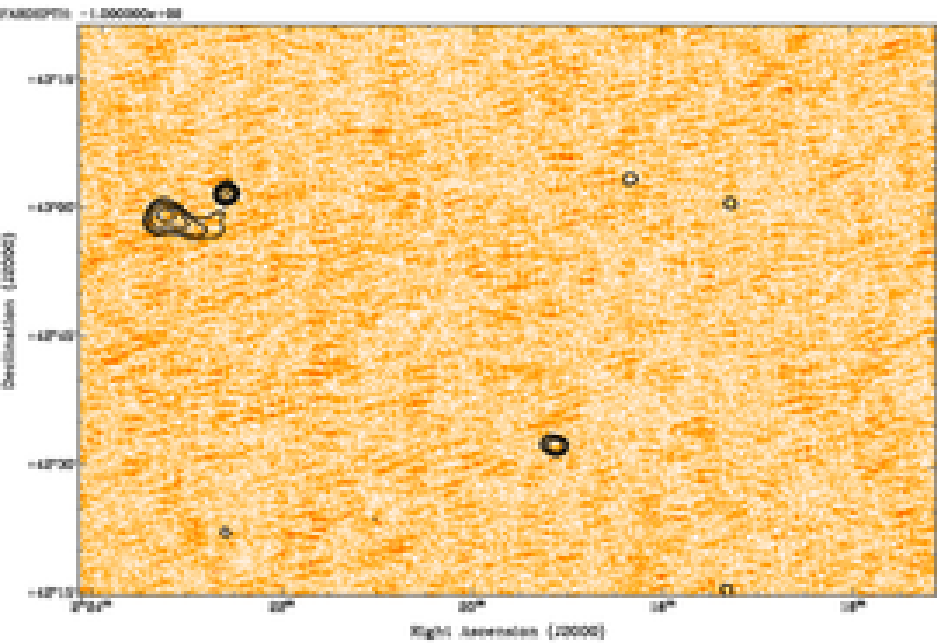}\hfill{}
\epsfxsize=0.49\textwidth \epsfbox{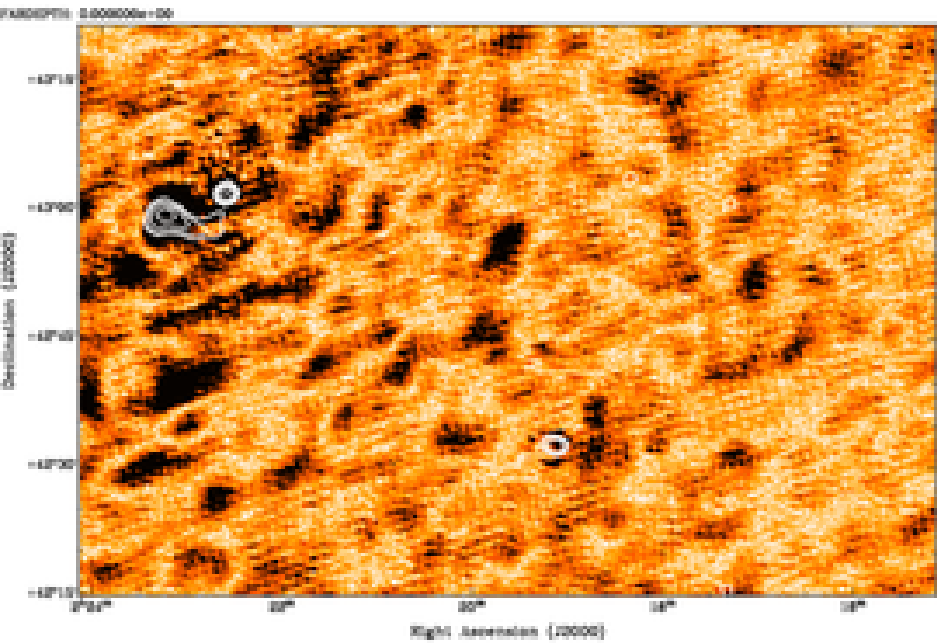}
\epsfxsize=0.49\textwidth \epsfbox{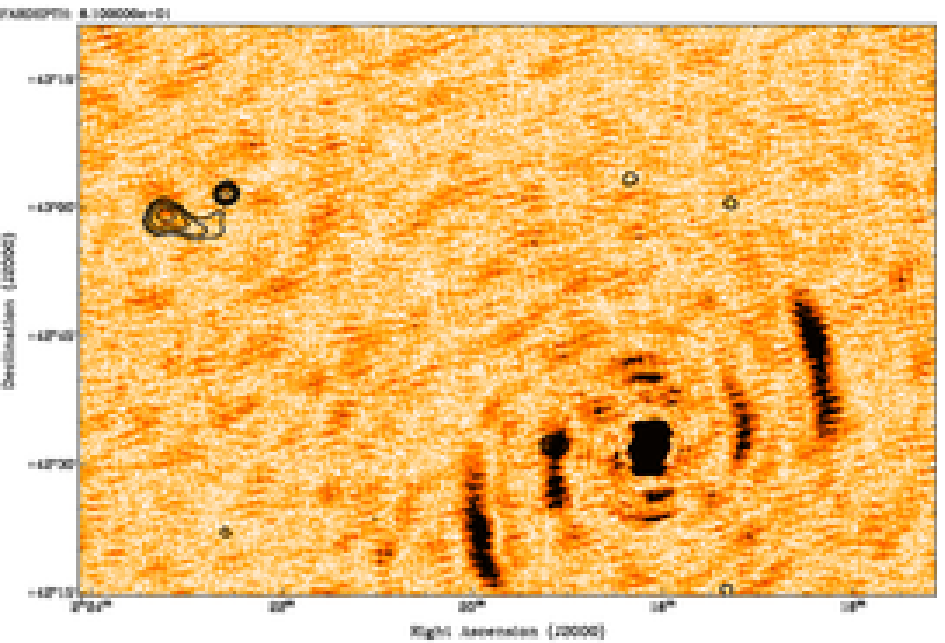}\hfill{}
\epsfxsize=0.49\textwidth \epsfbox{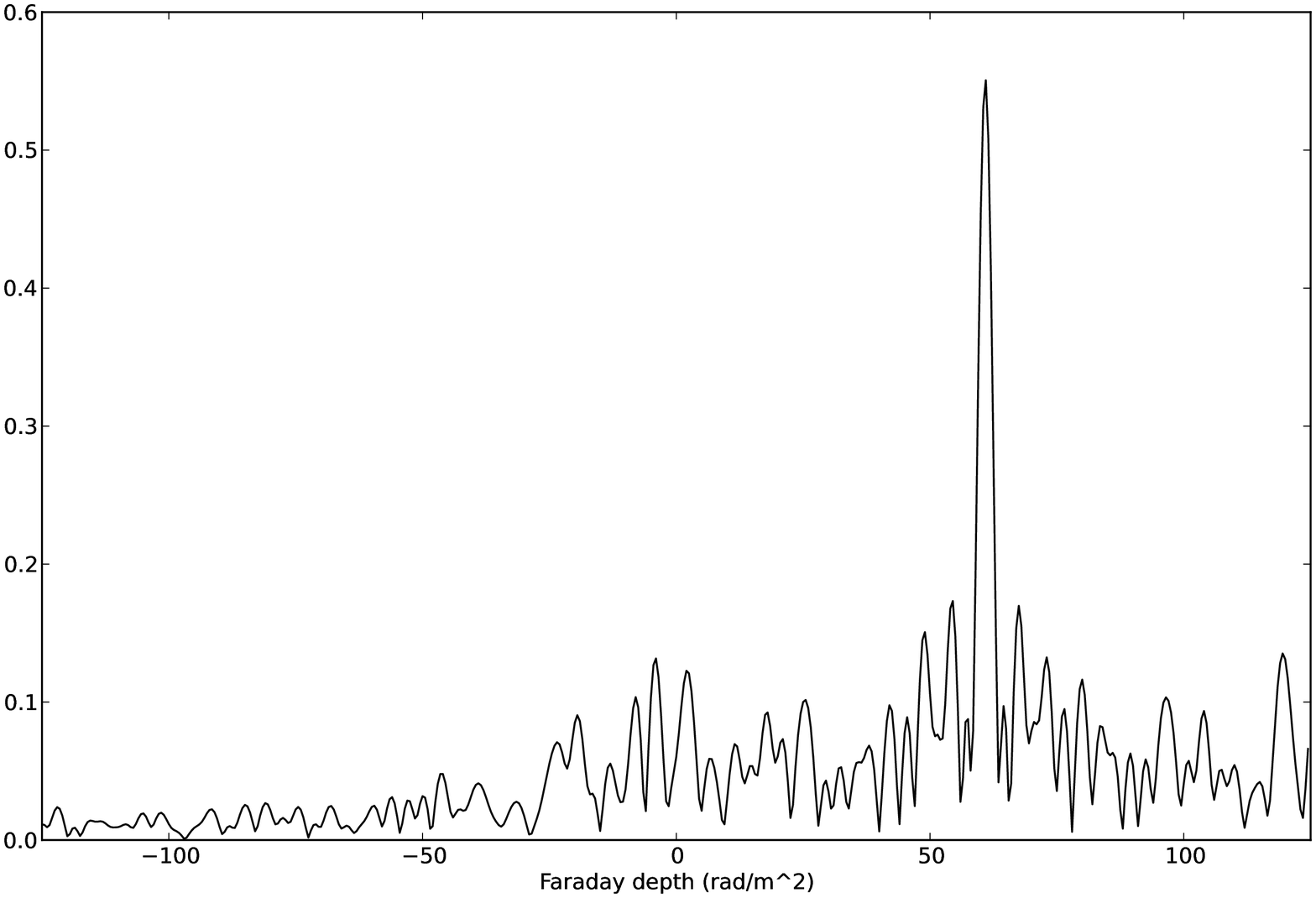}
\caption{RM Synthesis results from LOFAR observations of pulsar J0218+4232. Three frames are shown from the RM Synthesis cube, at Faraday depths of -100~rad~m$^{-2}$ (top left), 0~rad~m$^{-2}$ (top right), and +61~rad~m$^{-2}$ (bottom left). All three frames are displayed using the same colormap. A total intensity image (at a frequency of 138 MHz) is shown in contours, which start at 300~mJy~beam$^{-1}$ and increase by factors of two. The pulsar itself is seen in the frame at +61~rad~m$^{-2}$ (to the southwest), but not in the total intensity contours. The Faraday dispersion function of the pulsar is shown in the bottom right panel, illustrating that the polarized flux peaks at a Faraday depth of about +61~rad~m$^{-2}$. The incorrect sign of the Faraday depth is caused by a software mismatch in coordinate definitions.}
\end{figure}

\section{Prospects}

LOFAR has been producing excellent quality interferometric data since 2009. The commissioning period is proceeding well, and many of the difficult calibration issues intrinsic to low frequency radio interferometers, and low frequency aperture arrays in particular, have been addressed. As the commissioning period continues, we will begin our first large-scale survey of the sky, which is primarily designed to fill the calibration database, but will also provide high-quality source catalogs over a broad frequency range and the bulk of the northern sky.

After the commissioning period concludes, a combination of KSP observations and open-skies time will begin. Thanks to the strong emphasis on short baselines in the array design, combined with the excellent sensitivity of the telescope (especially in the HBA), LOFAR will be excellently suited to performing high fidelity observations of diffuse synchrotron emission in a range of astrophysical environments.

\section*{Acknowledgments}
LOFAR, the Low Frequency Array designed and constructed by ASTRON, has facilities in several countries, that are owned by various parties (each with their own funding sources), and that are collectively operated by the International LOFAR Telescope (ILT) foundation under a joint scientific policy.

The results presented here are the culmination of a great deal of work done by a large number of people. We gratefully acknowledge the engineers who have designed and built the array, as well as the large group of commissioners who have taken part in LOFAR ``busy weeks,'' where the bulk of the progress in imaging LOFAR data has taken place. C.~F. acknowledges financial support by the Agence Nationale de la Recherche through grant ANR-09-JCJC-0001-01. 


\label{lastpage}
\end{document}